\documentclass{article}
\newcommand{\bfr}{\begin{flushright}}
\newcommand{\efr}{\end{flushright}}
 
\begin{document}
\title{Magnetic Moment of Electrons near
Cosmic Strings
}
\author{Takuya Maki\\
Department of Physics, Tokyo Metropolitan University\\
Minami-osawa, Hachioji-shi, Tokyo 192-03, Japan\\
Kiyoshi
Shiraishi\\
Akita Junior College,\\
Shimokitade-sakura, Akita-shi, Akita 010, Japan
}
\date{Int. J. Mod. Phys. {\bf A9}, No. 11 (1994) pp. 1787--1795 }
\maketitle
\begin{abstract}
We study the effect of background geometry generated by a thin cosmic
string on the anomalous magnetic moment of the electron. We find that
the magnitude of the quantum correction to the magnetic moment depends
on the distance from the cosmic string as well as on the deficit angle.
\end{abstract}

\section{Introduction}
Topological defects will arise in some of the models with spontaneous
symmetry breakdown in field theory. Stringlike objects, which are
generated by the breakdown of U(1) symmetry, have the possibility of
providing the seeds for galaxies in the early universe.\cite{1,2,3}
The cosmic strings are expected to have large mass density and very
thin width. The space-time geometry around an infinitely stretching
straight string has a peculiar property. Suppose that the idealized
cosmic string lies with energy density (per unit length) $\mu$ on the
$z$ axis. The metric of the space-time is found to be \cite{1}
\begin{equation}
ds^2=-dt^2+dr^2+\frac{r^2}{\nu^2}d\theta^2+dz^2\,,
\label{1.1}
\end{equation}
where $\nu^{-1}=1-4G\mu < 1$, and $r$ and $\theta$ are the polar
coordinates of $x$-$y$ plane. If we use a new coordinate,
\begin{equation}
\theta'\equiv\frac{\theta}{\nu}\,,
\label{1.2}
\end{equation}
the metric reduces to the flat space-time except for the deficit in the
azimuthal angle because $\theta'$ takes a value from zero to $2\pi/\nu$.
Thus one can say that the space has a conical singularity at the
location of the idealized cosmic string.

Many authors have studied the quantum field theory around the conical
space.\cite{4} In such studies, the explicit calculation of quantum
effects is carried out owing to the local flatness of the space-time.
They have the same origin as the Casimir force \cite{5} between two
conducting plates.

The quantum electromagnetics near the conducting plates have also been
explored by many authors, including Barton et al. (see Ref.~\cite{6}
and references therein). The measurement of the $g$ factor has been
planned, though there is difficulty in obtaining sufficient precision
for the apparatus.

The cosmic strings, if associated with the GUT scale, are so thin that we can
regard them as ideal one-dimensional objects. Therefore we can
investigate quantum electromagnetism around cosmic strings in a similar
way. The analysis may reveal various types of quantum effects near
cosmic strings, which have not yet been calculated.

In the present paper, we calculate the anomalous magnetic moment of the
electron near a straight cosmic string with infinitesimal width. In
other words, we study quantum mechanics of an electron in the
background space-time expressed by the metric (\ref{1.1}). For
simplicity we consider the external magnetic field parallel to the
cosmic string which lies on the $z$ axis.

We must consider Ref.~\cite{6} when estimating the contribution to the
magnetic moment correction, though the manipulation related to the
regularization is quite different. The organization of this paper will
inevitably resemble the intensive work Ref.~\cite{6}.

Section~2 shows the preparation for the quantum-mechanical
calculations. The mode expansion of the electromagnetic field in the
conical spacetime is explicitly given. The renormalized value for the
mode summation is calculated. We show the Hamiltonian in the
nonrelativistic limit. In Sec.~3, we consider the second order
perturbation to the Hamiltonian for a ``fixed electron'' near a cosmic
string, which comes from the direct coupling term between the magnetic
moment of the electron and the magnetic field. Similar calculations of
the other contributions to the anomalous magnetic moment of the
electron near the cosmic string are performed in Sec.~4. Section~5 is
devoted to a summary and discussion.

\section{Preliminary: Mode Expansion of the Maxwell Field
in the Conical Space and the Hamiltonian}
We take the Coulomb gauge on the vector potential. Then ${\bf
E}=-\dot{\bf A}$ and
${\bf B}={\bf\nabla}\times{\bf A}$. 

The normal mode expansion in the
conical spacotime (\ref{1.1}) is written in the form 
\begin{equation}
A_i\approx\int_0^\infty dl\,l\int_{-\infty}^\infty
dk\frac{1}{\sqrt{2\omega}}\sum_n\sum_{s=1,2}a^sA_i^s(r,\theta)e^{ikz-i\omega
t}+h.c.\,,
\label{2.1}
\end{equation}
where the mode
functions $A_i^s$ are given by the sets
\begin{equation}
A_z^1=\left\{i\frac{1}{\omega}J_{\nu n}(lr)\sin n\theta\,,\quad
i\frac{1}{\omega}J_{\nu n}(lr)\cos n\theta
\right\}\,,
\label{2.2}
\end{equation}
\begin{eqnarray}
A_r^1&=&\left\{\frac{k}{2\omega}[J_{\nu n+1}(lr)-J_{\nu n-1}(lr)]\sin
n\theta\,,\right.\nonumber \\
& &\left.\frac{k}{2\omega}[J_{\nu n+1}(lr)-J_{\nu
n-1}(lr)]\cos n\theta
\right\}\,,
\label{2.3}
\end{eqnarray}
\begin{eqnarray}
A_\theta^1&=&\left\{-\frac{kr}{2\nu\omega}[J_{\nu n+1}(lr)+J_{\nu
n-1}(lr)]\cos n\theta\,,\right.\nonumber \\
& &\left.\frac{kr}{2\nu\omega}[J_{\nu n+1}(lr)+J_{\nu
n-1}(lr)]\sin n\theta
\right\}\,,
\label{2.4}
\end{eqnarray}
\begin{equation}
A_z^2=0\,,
\label{2.5}
\end{equation}
\begin{eqnarray}
A_r^2&=&\left\{\frac{1}{2}[J_{\nu n+1}(lr)+J_{\nu n-1}(lr)]\sin
n\theta\,,\right.\nonumber \\
& &\left.\frac{1}{2}[J_{\nu n+1}(lr)+J_{\nu
n-1}(lr)]\cos n\theta
\right\}\,,
\label{2.6}
\end{eqnarray}
\begin{eqnarray}
A_\theta^2&=&\left\{-\frac{1}{2}[J_{\nu n+1}(lr)-J_{\nu
n-1}(lr)]\cos n\theta\,,\right.\nonumber \\
& &\left.\frac{1}{2}[J_{\nu n+1}(lr)-J_{\nu
n-1}(lr)]\sin n\theta
\right\}\,,
\label{2.7}
\end{eqnarray}
$J_m(z)$ being the  Bessel function and $\omega=\sqrt{k^2+l^2}$.

We briefly  denote (\ref{2.1}) as
\begin{equation}
A_i=\sum_\lambda a_\lambda \bar{A}_{i\lambda}(r,\theta,z)e^{-i\omega
t}+h.c.\, ,
\label{2.8}
\end{equation}
where the single label $\lambda$ includes  $k, l,  n$  and  $s$,  and 
the $z$ dependence and the appropriate normalization are included in 
$\bar{A}_{i\lambda}$.
Accordingly, we can write
\begin{equation}
{\bf B}={\bf\nabla}\times{\bf A}=\sum_\lambda a_\lambda
\bar{\bf B}_{\lambda}(r,\theta)e^{-i\omega t}+h.c.\, ,
\label{2.9}
\end{equation}
etc. Upon the second quantization, $a$ and $a^\dagger$ are replaced by
the annihilation and creation operators. The calculation for
perturbative corrections can be written in simple forms by using the
notation.

The notation can be used for the ``inner product'' of the fields, which
is substantial in our calculation. We can express the vacuum
expectation value of the bilinear of quantum fields by using the mode
summation.\cite{7} For example, the unrenormalized expression for the
expectation value for the product of the vector field is written by
\begin{equation}
\langle A_zA^z\rangle=\sum_\lambda
\left|
\bar{A}_{z\lambda}\right|^2=\frac{\nu}{\pi}\sum_n\int_0^\infty
dl\,l\int_{-\infty}^\infty\frac{dk}{2\pi}\frac{1}{2\omega}
\frac{l^2}{\omega^2}J_{\nu n}^2(lr)\,.
\label{2.10}
\end{equation}

Similarly, one can obtain the expression
\begin{eqnarray}
& &\langle A_rA^r+A_\theta A^\theta\rangle=\sum_\lambda
(\bar{A}_{r\lambda}{\bar{A}^r_{\lambda}}{}^*+
\bar{A}_{\theta\lambda}{\bar{A}^\theta_{\lambda}}{}^*)\nonumber \\
& &\qquad=\frac{\nu}{\pi}\sum_n\int_0^\infty
dl\,l\int_{-\infty}^\infty\frac{dk}{2\pi}\frac{1}{2\omega}
\frac{k^2+\omega^2}{2\omega^2}[J_{\nu n+1}^2(lr)+
J_{\nu n-1}^2(lr)]\,.
\label{2.11}
\end{eqnarray}

The regularization for these quantities is rather simple in our case. We can
use Smith's method of regularization (see his paper in Ref.~\cite{4}).
His method is essentially equal to the point-splitting; one can consider
the small separation limit after replacing the arguments in the two
Bessel functions with $lr$ and $lr'$. An example of such calculation is
given in the Appendix. Here we write only the renormalized results:
\begin{equation}
\left\{\sum_\lambda\left|
\bar{A}_{z\lambda}\right|^2\right\}^{(R)}=\frac{\nu^2-1}{48\pi^2r^2}\,,
\label{2.12}
\end{equation}
\begin{equation}
\left\{\sum_\lambda(\bar{A}_{r\lambda}{\bar{A}^r_{\lambda}}{}^*+
\bar{A}_{\theta\lambda}{\bar{A}^\theta_{\lambda}}{}^*)\right\}^{(R)}=
\frac{\nu^2-1}{48\pi^2r^2}\,.
\label{2.13}
\end{equation}

We can also regularize the expression which appears in the perturbative
calculation. For later use, we calculate the quantities
\begin{equation}
\sum_\lambda\frac{1}{\omega_\lambda^2}\left|
\bar{B}_{z\lambda}\right|^2=\frac{\nu}{\pi}\sum_n\int_0^\infty
dl\,l\int_{-\infty}^\infty\frac{dk}{2\pi}\frac{1}{2\omega}
\frac{l^2}{\omega^2}J_{\nu n}^2(lr)\,,
\label{2.14}
\end{equation}
\begin{eqnarray}
& &\sum_\lambda\frac{1}{\omega_\lambda^2}
(\bar{B}_{r\lambda}{\bar{B}^r_{\lambda}}{}^*+
\bar{B}_{\theta\lambda}{\bar{B}^\theta_{\lambda}}{}^*)\nonumber \\
& &\qquad=\frac{\nu}{\pi}\sum_n\int_0^\infty
dl\,l\int_{-\infty}^\infty\frac{dk}{2\pi}\frac{1}{2\omega}
\frac{k^2+\omega^2}{2\omega^2}[J_{\nu n+1}^2(lr)+
J_{\nu n-1}^2(lr)]\,.
\label{2.15}
\end{eqnarray}

These mathematical expressions coincide with the previous ones for
vector fields.
Thus we immediately get the renormalized values:
\begin{equation}
\left\{\sum_\lambda\frac{1}{\omega_\lambda^2}\left|
\bar{B}_{z\lambda}\right|^2\right\}^{(R)}=\frac{\nu^2-1}{48\pi^2r^2}\,,
\label{2.16}
\end{equation}
\begin{equation}
\left\{\sum_\lambda\frac{1}{\omega_\lambda^2}
(\bar{B}_{r\lambda}{\bar{B}^r_{\lambda}}{}^*+
\bar{B}_{\theta\lambda}{\bar{B}^\theta_{\lambda}}{}^*)\right\}^{(R)}=
\frac{\nu^2-1}{48\pi^2r^2}\,.
\label{2.17}
\end{equation}

Now we turn to the Hamiltonian. Note that it takes the same form in the empty
space if we use the coordinate $\theta'$ [see (\ref{1.2})].
Consequently, we have only to take the effect of the wedge angle into
the mode sum calculations.

For the external (classical) magnetic field, we consider a uniform
magnetic field parallel to the cosmic string. We take the $z$ axis in
the direction
\begin{equation}
{\bf B}_0=B_0\hat{z}\,.
\label{2.18}
\end{equation}

Hereafter we separate the contribution of the external field 
from the vector potential:  ${\bf A}\rightarrow{\bf A}+{\bf A}_0$,
where ${\bf B}_0={\bf\nabla}\times{\bf A}_0$.

The mass and charge of the electron are denoted by $m$ and $e$,
respectively.

The  Hamiltonian in the nonrelativistic limit is given by \cite{6}
\begin{eqnarray}
H&=&H_{rad}+H_0-\frac{e}{2m}{\bf\sigma}\cdot{\bf B}-\frac{e}{m}
{\bf A}\cdot{\bf\pi}_0+\frac{e}{8m^2}{\bf\sigma}
\cdot({\bf\pi}_0\times{\bf E}-{\bf E}\times{\bf\pi}_0)\nonumber
\\ & &+\frac{e^2}{2m}{\bf A}^2+\frac{e^3}{4m^3}{\bf
A}^2{\bf\sigma}\cdot{\bf B}_0+\cdots\nonumber
\\ &=&H_{rad}+H_0+H_M+H_E+H_{SO}+H_2+H_3+\cdots\,, 
\label{2.19}
\end{eqnarray}
where $H_{rad}$ is the Hamiltonian for the Maxwell fields and
\begin{equation}
H_0=\frac{1}{2m}({\bf\pi}_0^2-e{\bf\sigma}\cdot{\bf B}_0)^2\,,
\label{2.20}
\end{equation}
with
\begin{equation}
{\bf\pi}_0=-i{\bf\nabla}-e{\bf A}_0\,.
\label{2.21}
\end{equation}
Here we omit the image energy (proportional to $e^2/r$), which is
irrelevant to the present calculation for the anomalous magnetic moment.

We will calculate the correction to the magnetic moment in the case of
$B_0 r\ll 1$, for we wish to know the leading order. Then we do not have
to worry about the other higher order terms in the expansion of the
Hamiltonian.

\section{Perturbative Calculation for a ``Fixed Electron''}
In this section, we treat an electron of fixed position, which is
abbreviated to a ``fixed electron.''\cite{5} We adopt the Hamiltonian
in the limit of infinite mass for the electron here. The treatment of
this quantum-mechanical correction to the magnetic moment is simple but
will be found to be short. The other contribution from quantum
``induction'' of the electromagnetic field will be treated in Sec.~4.

The spin of the electron is directly coupled to the external magnetic
field. The interaction Hamiltonian we examine in this section is simply
written as
\begin{eqnarray}
H_I&=&-{\bf\mu}\cdot{\bf B}_0-{\bf\mu}\cdot{\bf B}+H_{rad}\nonumber \\
&=&-\frac{e}{2m}{\bf\sigma}\cdot{\bf
B}_0-\frac{e}{2m}{\bf\sigma}\cdot\sum_\lambda(a_\lambda\bar{B}_\lambda
+a^\dagger_\lambda\bar{B}{}^*_\lambda)+\sum_\lambda\omega_\lambda
a^\dagger_\lambda a_\lambda\,,
\label{3.1}
\end{eqnarray}
where ${\bf\sigma}/2$ is the spin of the electron ($J=1/2$), and $\mu$
is the magnetic moment, $\mu=(e/m)({\bf\sigma}/2)$. By a ``fixed
electron'', we mean that only the interaction of the magnetic moment
and the magnetic field is taken into consideration.

The unrenormalized (second order) perturbation is given by \cite{5}
\begin{equation}
H_{eff}=-\sum_\lambda\sum_{M'=\pm1/2}\left|\left<M'\left|
\frac{e}{2m}{\bf\sigma}\cdot\bar{\bf B}_\lambda
\right|M\right>\right|^2
\frac{P}{\omega_\lambda+\left|\frac{e}{m}\right|B_0(M'-M)}\,,
\label{3.2}
\end{equation}
where $P$ means the principal value. Using the notation introduced in
Sec.~2, one can reduce (\ref{3.2}) to
\begin{eqnarray}
& &H_{eff}=-\frac{e^2}{m^2}\sum_\lambda\nonumber \\& &
\left\{\frac{M^2}{\omega_\lambda}
|\bar{\bf B}_{z\lambda}|^2+\frac{1}{4}\left[
\frac{\frac{1}{2}-M}{\omega_\lambda+\left|\frac{e}{m}\right|B_0}+
\frac{\left(\frac{1}{2}+M\right)P}{\omega_\lambda-\left|\frac{e}{m}
\right|B_0}\right](\bar{B}_{r\lambda}{\bar{B}^r_\lambda}{}^*+
\bar{B}_{\theta\lambda}{\bar{B}^\theta_\lambda}{}^*)
\right\}\,.
\label{3.3}
\end{eqnarray}

The leading contribution to the anomalous magnetic moment is a linear
part in $MB_0$ if we expand (\ref{3.3}) in the weak field limit. We find
that
\begin{equation}
-\delta{\bf\mu}\cdot{\bf B}_0=\frac{e^3}{4m^3}\sum_\lambda
\frac{2MB_0}{\omega_\lambda^2}(\bar{B}_{r\lambda}{\bar{B}^r_\lambda}{}^*+
\bar{B}_{\theta\lambda}{\bar{B}^\theta_\lambda}{}^*)\,.
\label{3.4}
\end{equation}
Using the renormalized value (\ref{2.17}), we obtain
\begin{equation}
\left(\frac{\delta\mu}{\mu}\right)_{fixed}=
-\frac{(\nu^2-1)e^2}{96\pi^2m^2r^2}\,.
\label{3.5}
\end{equation}

The correction to the magnetic moment of the same order of magnitude comes
from the other term in the Hamiltonian (\ref{2.19}). We study them in
the next section.

\section{More Perturbative Corrections}
In this section, we consider other contributions to the anomalous
magnetic moment of the electron near a cosmic string. By naive
consideration, the result of the previous section is satisfactory in
the nonrelativistic system. It was pointed out \cite{5} that the other
terms in the Hamiltonian, which are derived from the relativistic
theory, bring about additional contributions to the anomalous magnetic
moment of the same order as the previous result. They can be
interpreted as the effect of quantum ``electromagnetic induction.''

The first contribution we consider is the first order perturbation due
to $H_3$ in (\ref{2.19}). The effective Hamiltonian which comes from
this perturbation is given by
\begin{equation}
H_{eff}(3)=\frac{e^3}{4m^3}\left(
\sum_\lambda|\bar{\bf A}_\lambda|^2\right){\bf\sigma}\cdot{\bf B}_0\,.
\label{4.1}
\end{equation}
Using the renormalized values (\ref{2.12}) and (\ref{2.13}), we get
\begin{equation}
H_{eff}^{(R)}(3)=\frac{(\nu^2-1)e^3}{96\pi^2m^3r^2}{\bf\sigma}\cdot{\bf
B}_0\,.
\label{4.2}
\end{equation}

The next contribution originates from the second order perturbation due
to $H_E$ and $H_{SO}$. After some manipulation, we get the effective
Hamiltonian in the presence of the parallel magnetic field with the
cosmic string:
\begin{equation}
H_{eff}(E\cdot SO)=\frac{e^3}{4m^3}\left[
\sum_\lambda(\bar{A}_{r\lambda}{\bar{A}^r_\lambda}{}^*+
\bar{A}_{\theta\lambda}{\bar{A}^\theta_\lambda}{}^*)\right]{\sigma}_z
\cdot{B}_0\,.
\label{4.3}
\end{equation}
Applying the result (\ref{2.13}), we obtain
\begin{equation}
H_{eff}^{(R)}(E\cdot
SO)=\frac{(\nu^2-1)e^3}{192\pi^2m^3r^2}{\sigma}_z\cdot{B}_0\,.
\label{4.4}
\end{equation}

Finally, we consider the second order perturbation due to $H_E$ and
$H_M$ in (\ref{2.19}).
After lengthy calculation, we find that the effective Hamiltonian
includes the term
\begin{equation}
-\frac{e^3}{2m^3}\left[
\sum_\lambda\frac{1}{\omega_\lambda^2}|\bar{
B}_\lambda^z|^2\right]{\sigma}_z\cdot{B}_0
\label{4.5}
\end{equation}
if the external magnetic field is parallel to the $z$ axis,
on which the cosmic string lies. The substitution of the renormalized
value (\ref{2.16}) yields
\begin{equation}
-\frac{(\nu^2-1)e^3}{96\pi^2m^3r^2}{\sigma}_z\cdot{B}_0\,.
\label{4.6}
\end{equation}

The three additional contributions are of the same order as the result
obtained in the previous section.

Summing over all the contributions to the magnetic moment (\ref{3.5}),
(\ref{4.2}), (\ref{4.4}) and (\ref{4.6}), we find that the correction to
the magnetic moment is given by
\begin{equation}
\frac{\delta\mu}{\mu}=-\frac{(\nu^2-1)e^2}{48\pi^2m^2r^2}\,.
\label{4.7}
\end{equation}

\section{Summary and Discussion}
According to the result, the actual order of the magnitude is given by
\cite{5}
\begin{equation}
\frac{\delta\mu}{\mu}\approx\frac{(\nu^2-1)e^2}{m^2r^2}=1.09
\times10^{-23}\times(\nu^2-1)\times(r/{\rm cm})^{-2}\,.
\label{5.1}
\end{equation}

Since the width of cosmic strings associated with the GUT scale is about
$10^{-28}$ cm and $\nu-1\approx 10^{-6}$, it is possible for electrons
to approach so closely that the correction to the anomalous magnetic
moment becomes large. Thus, if one can prepare a static  cosmic 
string in his laboratory, he could measure the shifted value for
the anomalous magnetic moment of electrons. 

In astronomical observation, however, we do not expect to find  any 
effect even if cosmic strings exist in our universe, because:

(1) The strong magnetic field and the high density ionized region can
hardly overlap each other in the galactic space. [There is a little
hope in the case where the cosmic string is located in the vicinity of
a neutron star (with an ac companied star).]

(2) Cosmic strings are expected to be moving near the speed of light,
so that other effects may be quite large and eliminate the effect
considered in this paper. Dynamical effects on classical and quantum
electromagnetism must be studied. 

The calculation of the cyclotron
frequency shift near a cosmic string is more difficult in our case than
that near conducting plates.\cite{6} This task is left for future works.

\section*{Acknowledgment}
The authors would like to thank Dr. Y. Kojima for useful comments.

\section*{Appendix}
We give an illustrative calculation for the regularization of the mode
summation. We consider here the expression
\begin{equation}
\sum_\lambda
\left|
\bar{A}_{z\lambda}\right|^2=\frac{\nu}{\pi}\sum_n\int_0^\infty
dl\,l\int_{-\infty}^\infty\frac{dk}{2\pi}\frac{1}{2\omega}
\frac{l^2}{\omega^2}J_{\nu n}^2(lr)\,.
\label{A.1}
\end{equation}
We regard this as the following quantity in the limit of
$\zeta\rightarrow 0$ and $r'\rightarrow r$:
\begin{equation}
\frac{\nu}{\pi}\frac{1}{2}\sum_n\int_0^\infty
dl\,l\int_{-\infty}^\infty\frac{dk}{2\pi}\frac{1}{2\omega}
\frac{l^2}{\omega^2}J_{\nu n}(lr)J_{\nu n}(lr')e^{ik\zeta}\,,
\label{A.2}
\end{equation}
where $\zeta$ is the difference of the $z$ coordinates of two separated
points. 

By  using  the formula on  the modified Bessel  functions, 
which  can  be found in Smith's paper (Ref.~\cite{3}), we reduce
(\ref{A.2})  to
\begin{equation}
\frac{1}{4\pi^2rr'}\frac{\nu}{\sinh\frac{\nu u}{2}}\frac{\cosh
u-1}{\sinh^2u}\left(\frac{\cosh u}{\sinh u}\cosh\frac{\nu
u}{2}+\frac{\nu}{2\sinh\frac{\nu u}{2}}\right)\,,
\label{A.3}
\end{equation}
where
\begin{equation}
\cosh u\equiv\frac{r^2+r'^2+\zeta^2}{2rr'}\,.
\label{A.4}
\end{equation}
For small $u$, this can be expanded as
\begin{equation}
\frac{1}{2\pi^2rr'}\frac{1}{u^2}\left(1+\frac{1}{12}u^2+
\frac{\nu^2u^2}{24}+\cdots\right)\,.
\label{A.5}
\end{equation}
We can subtract the result for the empty space --- which can be obtained
by setting $\nu=1$ --- from the expression (\ref{A.5}). After the
subtraction, we can take the limit $u\rightarrow 0$ and $r'\rightarrow
r$. This operation leads to
\begin{equation}
\left\{\sum_\lambda
\left|
\bar{A}_{z\lambda}\right|^2\right\}^{(R)}=\frac{\nu^2-1}{48\pi^2r^2}\,.
\label{A.6}
\end{equation}

We can get the normalized quantities of the other types by similar methods.



\begin{thebibliography}{99}
\bibitem{1} A. Vilenkin, Phys. Rep. {\bf 121} (1985) 263.

\bibitem{2} {\it Cosmic Strings: The Current Status}, eds. F. S.
Accetta and L. M. Krauss (World Scientific, Singapore, 1988).

\bibitem{3} {\it The Formation and Evolution of Cosmic Strings}, eds. G.
Gibbons, S. Hawking and T. Vachaspati (Cambridge University Press,
Cambridge, 1989).

\bibitem{4} V. P. Frolov and E. M. Serebriany, Phys. Rev. {\bf D35}
(1987) 3779; B. Linet, Phys. Rev. {\bf D35} (1987) 536; J. S. Dowker,
Phys. Rev. {\bf D36} (1987) 3742; J. S. Dowker, in Ref.~3, p. 251; A.
G. Smith, in Ref.~3, p. 263; A. Sarmiento and S. Hacyan, Phys. Rev.
{\bf D38} (1988) 1331; I. H. Russell and D. J. Toms, Class. Quantum
Grav. {\bf 6} (1989) 1343; D. Harari and V. Skarzhinsky, Phys. Lett.
{\bf B240} (1990) 322; J. Audretsch and A. Economou, Phys. Rev. {\bf
D44} (1991) 980; S. Hirenzaki and K. Shiraishi, Class. Quantum
Grav. {\bf 9} (1992) 2277.

\bibitem{5} H. B. G. Casimir, Proc. Kon. Ned. Akad. Wet. {\bf 51}
(1948) 793;
 G. Plunien, B. M\"uller
and W. Greiner, Phys. Rep. {\bf 134} (1986) 87; V. M. Mostepanenko and
N. N. Trunov, Sov. Phys. Usp. {\bf 31} (1988) 965.

\bibitem{6} G. Barton and N. S. J. Fawcett, Phys. Rep. {\bf 170} (1988)
1.
\bibitem{7} K. Shiraishi, J. Korean Phys. Soc. {\bf 25} (1992) 192.
\end{thebibliography}
\end{document}